\documentclass[final]{siamltex}
\usepackage{amsmath,amssymb,cite}
\usepackage{graphicx}
\usepackage{amsbsy}

\author{Tobias Sch{\"a}fer\thanks{
        Department of Mathematics, The College of Staten
        Island, City University of New York, New York
        ({\tt tobias@math.csi.cuny.edu}).}
        \and Richard O. Moore\thanks{
        Department of Mathematics,
        New Jersey Institute of Technology,
        Newark, New Jersey
        ({\tt rmoore@oak.njit.edu}).} }

\title{A Path Integral Method for Coarse-Graining Noise in
Stochastic Differential Equations with Multiple Time Scales}

\newcommand{\dav}{d_\mathrm{av}}
\newcommand{\bp}{\begin{pmatrix}}
\newcommand{\ep}{\end{pmatrix}}

\begin{document}

\bibliographystyle{unsrt}

\maketitle
\begin{abstract}
We present a new path integral method to analyze stochastically
perturbed ordinary differential equations with multiple time scales.
The objective of this method is to derive from the original system a
new stochastic differential equation describing the system's
evolution on slow time scales. For this purpose, we start from the
corresponding path integral representation of the stochastic system
and apply a multi-scale expansion to the associated path integral
kernel of the corresponding Lagrangian. As a concrete example, we
apply this expansion to a system that arises in the study of random
dispersion fluctuations in dispersion-managed fiber optic
communications. Moreover, we show that, for this particular example,
the new path integration method yields the same result at leading
order as an asymptotic expansion of the associated Fokker-Planck
equation.
\end{abstract}

\section{Introduction}

Physical phenomena exhibiting scale separation are ubiquitous in a
wide range of fields, most obviously in those permitting a
description via continuum mechanics such as fluid dynamics,
electromagnetism, and material science.  The ability to separate the
(macroscopic) evolution on slow time scales from the (microscopic)
evolution on fast time scales is key to understanding a system's
behavior or to performing numerical simulations in an efficient way.
Many mathematical techniques have been developed for explicitly
performing this separation of scales in the absence of stochastic
fluctuations, e.g. multi-scale expansions \cite{holmes:1995}, Lie
transform \cite{nayfeh:1973}, and renormalization group methods
\cite{chen-goldenfeld-etal:1994}. In a typical multi-scale analysis,
we start from a given system, which we write here as an ordinary
differential equation for a vector $x=x(t)$,
\begin{displaymath}
\frac{dx}{dt}=f(x,t,\epsilon)\;,
\end{displaymath}
where $\epsilon$ is a small parameter. Upon introducing, for
instance, two time scales $t_0=t$ and $t_1=\epsilon^{\gamma}t$ (the
actual power of $\gamma>0$ is selected by the problem), we derive
systematically a new ordinary differential equation
\begin{displaymath}
\frac{dX}{dt_1}=F(X,t_1)\;,
\end{displaymath}
describing the evolution of the system on the {\em slow} time scale
represented by $t_1$ through $X=X(t_1)$. In this article, we are
interested in the effect of stochastic perturbations on this
description over the coarse-grained (slow) scales.  Intuitively we
expect to derive from the microscopic SDE given by
\begin{displaymath}
\frac{dx}{dt}=f(x,t,\epsilon)+g(x,t,\epsilon)\xi(t)\,,
\end{displaymath}
where $\xi$ represents white noise, a new stochastic differential
equation of the form
\begin{displaymath}
\frac{dX}{dt_1}=F(X,t_1)+G(X,t_1)\Xi(t_1),
\end{displaymath}
describing the slow scale evolution, where $\Xi$ represents white noise on
the slow time scale $t_1$. Our intent is not to explore under
what circumstances or over what range of parameters such an approximation
is valid; rather, we present a
practical method based on path integrals that can be used for explicitly
calculating the coarse-grained stochastic differential equation from the original
system, implicitly assuming that such an approach is valid.

For this purpose, we now state the problem in a slightly more formal
way: Assume a system of stochastic ordinary differential (Langevin)
equations expressed by
\begin{equation}
\dot{x} = f(x,t;\epsilon) + \sqrt\epsilon g(x,t)\xi(t),\quad x(0)=a
\label{SODE}
\end{equation}
where $x\in \mathbb{R}^n$, $t\in \mathbb{R}^+$,
$f:[\mathbb{R}^n\times\mathbb{R^+}]\rightarrow\mathbb{R}^n$,
$g:[\mathbb{R}^n\times\mathbb{R^+}]\rightarrow\mathbb{R}^{n\times
m}$, and $f$ is assumed to be sufficiently differentiable in
$\epsilon$. The random term, $\xi(t)$, is assumed to be
$m$-dimensional delta-correlated white noise, i.e., with
\begin{equation}
<\xi_p(t)\xi_q(t')> = \delta(t-t')\delta_{pq}
\end{equation}
where $<\cdot>$ denotes the ensemble average, and
Gaussian-distributed with a probability density given by
\begin{equation}
p(\xi) = \frac1{(2\pi)^{m/2}}\exp(-\frac12~\xi\cdot\xi).
\end{equation}
We choose to take the Stratonovich interpretation of
Eqn.~\ref{SODE} given our interest in modeling physical
processes; however, this is simply a convention given the general
form of Eqn.~\ref{SODE}.  Further assumptions will be indicated as
they become necessary.

The goal is to obtain {\em another} stochastic ODE that captures
the evolution of $x$ on a slow scale when $0<\epsilon\ll 1$, i.e.,
to ``scale up" the noise from its microscopic representation to a
macroscopic noise that reflects the fast microscopic response of
$x$ in addition to the noise itself.  The ability to integrate a
macroscopic stochastic equation on a much slower time scale, where
the microscopic trajectories can be reconstructed approximately
from the macroscopic trajectories, is of obvious benefit from the
perspectives of asymptotic analysis and numerical generation of
the statistical properties of $x$.

For the reader who is less acquainted with the scaling of small
parameters in stochastic equations, we comment briefly on the factor
$\sqrt{\epsilon}$ in the problem (\ref{SODE}) in the context of
an expansion using multiple time scales.
A naive expansion of $x$ in Eqn.~\ref{SODE} in powers of $\epsilon$
($\sqrt\epsilon$) is quickly seen to become disordered when $t\sim
1/\epsilon$ ($t\sim 1/\sqrt\epsilon$). A standard method that is
widely used in the deterministic case \cite{holmes:1995} to address
this onset of secularity is the use of multiple scales in time, i.e.,
let
\begin{equation}
x = x^{(0)}(t_0,t_1,\dots) + \epsilon x^{(1)}(t_0,t_1,\dots) + \dots,
\label{MSexpansion}
\end{equation}
where $t_j = \epsilon^j t$.  Whereas the dependence of $x^{(0)}$,
$x^{(1)}$, etc., on the ``fast" time variable $t_0$ alone is
insufficient to prevent this expansion of $x$ from becoming
disordered, their dependence on the ``slow" time variables $t_1$,
etc., allows sufficient freedom to suppress secular growth up
to times of order $1/\epsilon$. Before proceeding, we have to justify the
particular expansion suggested in Eqn.~\ref{MSexpansion}. For
simplicity, we do this in a one-dimensional context; the argument
generalizes in an obvious way to higher dimensions. Were $\xi(t)$ a
{\em deterministic\/} forcing term, we would ordinarily expand $x$
and $t$ in powers of $\sqrt\epsilon$.  In this case, however,
$\xi(t)$ is formally the time-derivative of a Wiener process $W(t)$,
with Eqn.~\ref{SODE} more appropriately expressed in the form
\begin{equation}
dx = f(x,t;\epsilon)\,dt + \sqrt\epsilon g(x,t)\,dW,
\end{equation}
where $\langle W(t)W(t')\rangle = \min(t,t')$, i.e., with $\langle
dW^2\rangle \equiv \langle (W(dt)-W(0))^2\rangle= dt$.  As $x(t)$ is
a stochastic process, we wish to track all of its moments, as an
alternative to the direct evolution of the probability density
function via the Fokker-Planck equation (see Sec.~\ref{FPsec}).  In
a time step of $dt$, we have
\begin{eqnarray*}
\langle x(t+dt)-x(t)\rangle &=& \langle dx\rangle = \langle
f(x,t;\epsilon)\rangle \,dt,\quad\mbox{and}\\
\langle x^2(t+dt) - x^2(t)\rangle &=& 2\langle x\,dx\rangle +
\langle dx^2\rangle\\
&=& 2\langle xf(x,t;\epsilon)\rangle\,dt + \epsilon\langle g^2(x,t)\rangle
\,dt,
\end{eqnarray*}
so that all terms of ${\mathcal O}(\epsilon^{j/2})$ vanish for $j$
odd.  An expansion in powers of $\epsilon$ is therefore
self-consistent.  We also see from this consideration that the
influence of the stochastic driving term in Eqn.~\ref{SODE} is first
felt at ${\mathcal O}(\epsilon)$, suggesting that the term of
${\mathcal O}(\sqrt\epsilon\,dW)$ mimics a term of ${\mathcal
O}(\epsilon\,dt)$.

The paper is organized as follows: In Sec.~\ref{PathIntsec}, we
present the path integral method for deriving the slow stochastic
dynamics from the original system. In Sec.~\ref{Dispsec}, we apply
this method to a particular problem in fiber optics. In what
follows, we show that the new method, for the chosen example, yields
the same result as an asymptotic expansion of the associated
Fokker-Planck equation (Sec.~\ref{FPsec}). Conclusions are presented
in Sec.~\ref{Concsec}.

\section{Averaging of path integral kernels}
\label{PathIntsec}

Instead of considering the stochastic differential Eqn.~\ref{SODE}
directly we can consider the probability density $p(x,t)$ corresponding
to the stochastic process $x$ and write down a path integral representation
for $p(x,t)$\cite{feynman-hibbs:1965,chaichian-demichev:2001,arnold:2000}:
\begin{equation} \label{pathint_rep}
p(x,t) = \int_{{\mathcal{C}}(x,t|a,0)} {\mathcal{D}}x(\tau)
\;{\mathrm{e}}^{-\int_0^t\;L(x(\tau),\dot x(\tau),\tau)\;d\tau} =
\int_{{\mathcal{C}}(x,t|a,0)} {\mathcal{D}}x(\tau)
{\mathcal{P}}(x(\tau))
\end{equation}
with the Lagrangian $L$ given by
\begin{equation}
L = \frac{1}{2\epsilon}  \sum_{i=1}^{n} h_{ik}^2 \left(\dot x_k
-\left(f_k+\epsilon s \frac{\partial g_{kj}}{\partial
x_l}g_{lj}\right)\right)^2,
\end{equation}
written according to Einstein's summation notation (see Appendix).
 Here we set $s=1/2$ according to the Stratonovich interpretation of
Eqn.~\ref{SODE}. In cases where $h=g^{-1}$ does not exist, it is
necessary to consider a regularized path integral representation.  A
short derivation for both cases can be found in the appendix. In
Eqn.~\ref{pathint_rep}, we take all paths $\Gamma \in
{{\mathcal{C}}(x,t|a,0)}$ into account, that lead from the initial
point $(a,0)$ to the final point $(x,t)$. We want to develop a
multi-scale technique on the level of this path integral
representation and assume a periodicity on the fast time scale $t_0$
with a period $t^*$. Assume that the final time $t=(K+1)t^*$,
meaning that we propagate for $K+1$ periods and with $K$
intermediate times $t_{0k}=kt^*$. The fast time scale $t_0=t$
characterizes the system's evolution within one period, e.g. for a
time interval $[t_{0k},t_{0k}+t^*]$. The evolution on the time scale
$t_1=\epsilon t$ characterizes the slow evolution of the system. For
a given path $\Gamma \in {{\mathcal{C}}(x,t|a,0)}$, we take
$X_k=x_k=x(kt^*)$, $0\leq k\leq K$ as sample points of the process
$x$ at times $t_{1k} = \epsilon kt^*$ and we are looking for a
slowly varying process $X=X(t_1)$ as a continuum limit of this
sampled process $(X_k)$. We can write down the transition
probability $p(x,t)\equiv p(x,t|a,0)$ as a $K$-dimensional integral
\begin{eqnarray*}
p(x,t|a,0) &=& \int \left (\prod_{k=0}^K p(x_{k+1},(k+1)t^*|x_k,kt^*)\right)\,dx_1...dx_K \\
               &=& \int \left (\prod_{k=0}^K \bar p(X_{k+1},t_{1(k+1)}|X_k,t_{1k})\right)\,
               dX_1...dX_K
\end{eqnarray*}
with
\begin{equation}
\bar p(X_{k+1},t_{1(k+1)}|X_k,t_{1k})=p(x_{k+1},(k+1)t^*|x_k,kt^*).
\end{equation}
In order to use the path integral representation given in
Eqn.~\ref{pathint_rep}, we can write any path $\Gamma \in
{\mathcal{C}}(x,t|a,0)$ as
\begin{displaymath}
\Gamma = \sum_{k=0}^{K} \Gamma_k, \qquad \Gamma_k \in
{\mathcal{C}}(x_{k+1},(k+1)t^*|x_k,kt^*).
\end{displaymath}
The following Figure~\ref{fig:pathfig} illustrates this decomposition
of a path into polygons connecting the start- and endpoints of the spans
and the paths within one period.
\begin{figure}[htb]
\centering
\begin{minipage}{0.95\textwidth}
\begin{center}
\includegraphics[width=0.7\textwidth, angle=-90]{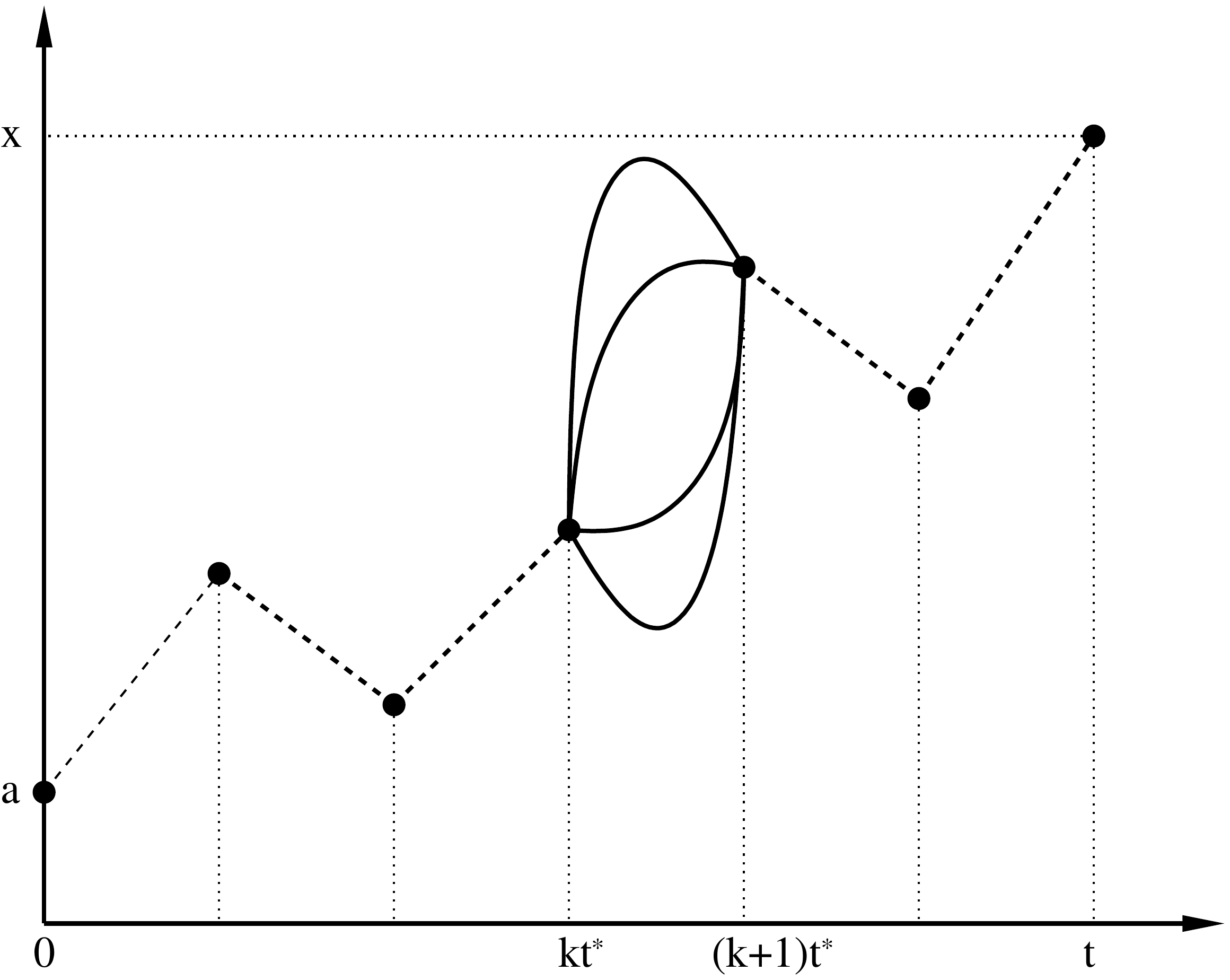}
\caption{{\small A path from $(a,0)$ to $(x,t)$ can be decomposed
into a polygon connecting the beginning and the endpoints of the
periods and paths within the periods. The short-time propagator
averages over all paths within one period. As an example, the paths
from time $kt^*$ to time $(k+1)t^*$} are drawn as solid lines.}
\label{fig:pathfig}
\end{center}
\end{minipage}
\end{figure}
The transition probability $p(x_{k+1},(k+1)t^*|x_k,kt^*)$ from the
beginning of a period to its end is then given by the path integral
\begin{equation} \label{fast_pathint}
p(x_{k+1},(k+1)t^*|x_k,kt^*) =
\int_{{\mathcal{C}}(x_{k+1},(k+1)t^*|x_k,kt^*)}
{\mathcal{P}}(x(\tau)){\mathcal{D}}x(\tau)
\end{equation}
and in the continuum limit for the slow scale $t_1$, the polygons
connecting $(a,0)$ with $(x,t)$ through $K$ intermediate points
become paths $X(t_1)$ and the probability distribution for the
polygons becomes the probability distribution
$\bar{\mathcal{P}}(X(t_1))$ of these paths
\begin{equation}
\lim_{K\rightarrow \infty,\; \delta\rightarrow 0} \left
(\prod_{k=1}^K \bar p(X_{k+1},t_{1(k+1)}|X_k,t_{1k})\right) =
\bar{\mathcal{P}}(X(t_1)).
\end{equation}
Thus we obtain the final representation of the transition probability
on the slow scales:
\begin{equation} \label{slow_pathint}
p(x,t) = \bar p(X,t_1|a,0) =
\int_{{\mathcal{C}}(X,t_1|a,0)}\bar{\mathcal{P}}(X(\tau_1)){\mathcal{D}}X(\tau_1)
\end{equation}
The last three equations are the mathematical formulation of the
intuitive approach: First, one considers the dynamics of the system on
the fast scales. The resulting path integral yields the propagator of
the system from the beginning of a period to the end of the
period. Combining these propagators leads to the probability of the
path on the slow scales. Therefore, as an abbreviation for combining
Eqn.~\ref{fast_pathint} and Eqn.~\ref{slow_pathint}, we can write the
transition probability $p$ as a hierarchy of path integrals:
\begin{equation} \label{formal_pathint}
p(x,t|a,0) = \int \left( \lim \prod \int {\mathcal{P}}(X(\tau_0)){\mathcal{D}}X(\tau_0)
                  \right) {\mathcal{D}}X(\tau_1)
\end{equation}
This method can be generalized in an obvious way: Introducing $J$
time scales $t_j = \epsilon^j t$, the product of the propagators on
the time scale $t_j$ leads to the probability of a path on the time
scale $t_{j+1}$. Using the same abbreviated notation as in
Eqn.~\ref{formal_pathint}, we can represent $p=p(x,t|a,0)$ as
\begin{equation} \label{pathint_hiearchy}
p = \int ... \left( \lim \prod \int \left( \lim \prod \int
{\mathcal{P}}(X(\tau_0)){\mathcal{D}}X(\tau_0) \right)
{\mathcal{D}}X(\tau_1) \right)... {\mathcal{D}}X(\tau_J)
\end{equation}
Once we have obtained the path integral representation on the slow scale
with an averaged kernel
\begin{equation}
\bar{\mathcal{P}}(X(\tau_1))={\mathrm{e}}^{-\int_0^{t_1}\bar L(X(\tau_1),\dot
X(\tau_1),\tau_1)d\tau_1}
\end{equation}
we can also convert back in order to find the corresponding stochastic
equation on the slow time scale.

\section{Dispersion-managed fiber-optic communications}
\label{Dispsec}

To demonstrate the application of the method described above,
we consider the propagation of an electromagnetic field through an
optical fiber with a periodic, piecewise-constant coefficient of
dispersion.  The model equation for this process is the
dispersion-managed nonlinear Schr\"odinger (DMNLS) equation given by
\begin{equation}
iE_z + d(z)E_{tt} + \epsilon|E^2|E = 0, \label{DMNLS}
\end{equation}
where $z$, $t$, $E(t,z)$, $d(z)$ are dimensionless quantities
representing distance down the fiber (the ``time-like" evolution
variable in signaling coordinates), time (the transverse variable
in signaling coordinates), the electric field envelope and the
dispersion coefficient.  The scaling parameter $0<\epsilon\ll 1$
reflects the typical case of a weak nonlinearity relative to the
dispersion strength at any point in the fiber.  Under typical
deterministic dispersion management, the fiber dispersion has a
small positive (anomalous) mean value and $\mathcal{O}(1)$ local
mean-zero value, i.e., $d(z) = \epsilon \dav + d_0(z)$, where, for
instance,
\begin{equation}
d_0(z) =
\begin{cases}
+\hat{d}    & 0\leq z<1/4\\
-\hat{d}    & 1/4\leq z<3/4\\
+\hat{d}    & 3/4\leq z<1\\
\end{cases},
\label{DMmap}
\end{equation}
and where $d_0(z)$ has unit periodicity.  Here, we consider the
case where $d(z)$ has an additional random component, giving $d(z)
= \epsilon\dav + d_0(z) + \sqrt{\epsilon D}\xi(z)$, where $\xi(z)$
is taken to be $\delta$-correlated white noise with unit strength,
i.e.,
\begin{equation}
<\xi(z)\xi(z')> = \delta(z-z').
\end{equation}
For more details on the derivation of this model from Maxwell's
equations, refer to \cite{newell-moloney:1992,agrawal:1995}.
It has already been
demonstrated~\cite{abdullaev-bronski-etal:2000,schaefer-moore-etal:2002}
that the effect of this random dispersion on single periodic solutions
(referred to as ``dispersion-managed solitons") of Eqn.~\ref{DMNLS}
can effectively be captured by the two-dimensional system of
stochastic ordinary differential equations given by
\begin{eqnarray*}
\frac{dT}{dz} & = & 4d(z)M\quad\mbox{and}\\
\frac{dM}{dz} & = & \frac{C_1d(z)}{4T^3} - \frac{\epsilon
C_2}{4T^2}, \label{TMeqns}
\end{eqnarray*}
where $T(z)$ and $M(z)$ represent the width and chirp (i.e., the
quadratic component of the phase) of the soliton.  For derivations
of this system using a variational principle, by applying the lens
transform followed by expansion in Gauss-Hermite modes, or using
moments with a closure condition, see
\cite{turitsyn:1998,lakoba-kaup:1998}.  Constants $C_1$ and $C_2$
depend on the particular shape assumed for the soliton. When
$\epsilon=0$, these equations simply reflect the linear
Schr\"odinger equation with mean-zero dispersion, and all initial
conditions, in particular those of the form
$\{T(0),M(0)\}=\{T_0,0\}$, yield periodic solutions. When $\epsilon$
is finite and $D=0$ (i.e., the dispersion is deterministic and
periodic), the nonlinearity and finite-mean dispersion ``select" a
member of this one-dimensional family of periodic solutions that
persists for $\epsilon>0$~\cite{cautaerts-maruta-etal:2000}. When
$\epsilon>0$ and $D=\mathcal{O}(1)$, the width and chirp of this
soliton undergo a random walk in addition to the ``breathing"
induced by the deterministic component of the dispersion map.  We
now seek to apply the methods discussed in the previous sections to
determine the properties of this random walk.

Before we proceed, it is useful to note that the action-angle
coordinates \cite{turitsyn-aceves-etal:1998} in the $\epsilon=0$ case
also simplify analysis for the finite-$\epsilon$ case.
Letting $\Omega = C_1/T^2+16M^2$ and $\beta=4TM$ gives
\begin{eqnarray}
\frac{d\Omega}{dz} & = & 3\epsilon\frac{\partial h}{\partial\beta}\quad\mbox{and}\nonumber\\
\frac{d\beta}{dz} & = & d_0(z)\Omega - \epsilon \frac{\partial
h}{\partial\Omega} \label{AAeqns}
\end{eqnarray}
where
\begin{equation} \label{tm_hamiltonian}
h(\Omega,\beta) =
\frac{2C_2\Omega^{3/2}}{3(C_1+\beta^2)^{1/2}}-\frac12
\dav\Omega^2.
\end{equation}
An initial condition in Eqns.~\ref{TMeqns} equal to $\{T_0,0\}$
corresponds to an initial condition in Eqns.~\ref{AAeqns} of
$\{\Omega_0,0\}$, where $\Omega_0 = C_1/T_0^2$.

\begin{figure}[htb]
\centering
\begin{minipage}{0.95\textwidth}
\begin{center}
\includegraphics[width=0.9\textwidth, angle=0]{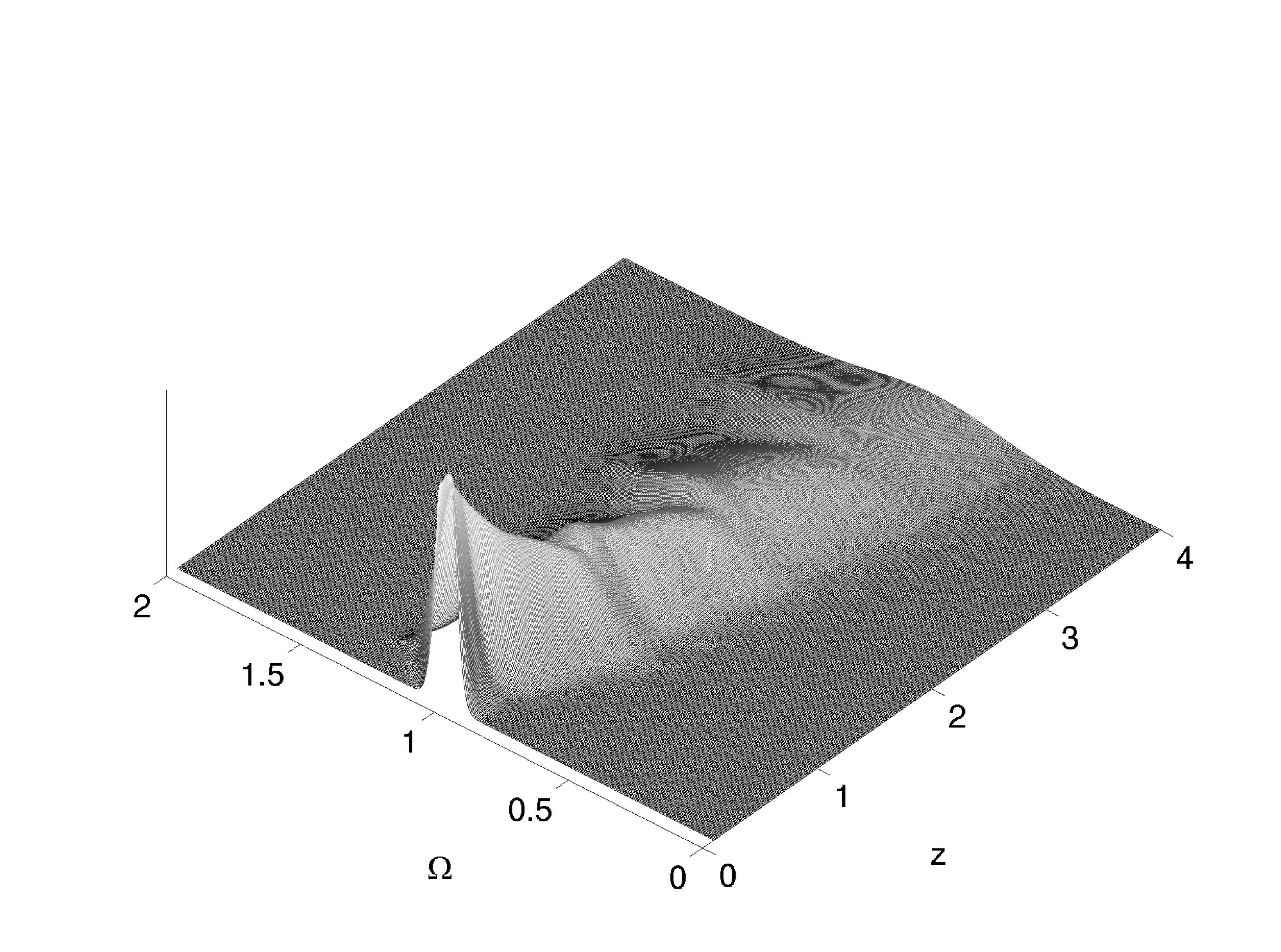}
\caption{{\small Evolution of the probability density of $\Omega$}}
\label{fig:Ommargin}
\end{center}
\end{minipage}
\end{figure}

\begin{figure}[htb]
\centering
\begin{minipage}{0.95\textwidth}
\begin{center}
\includegraphics[width=0.9\textwidth, angle=0]{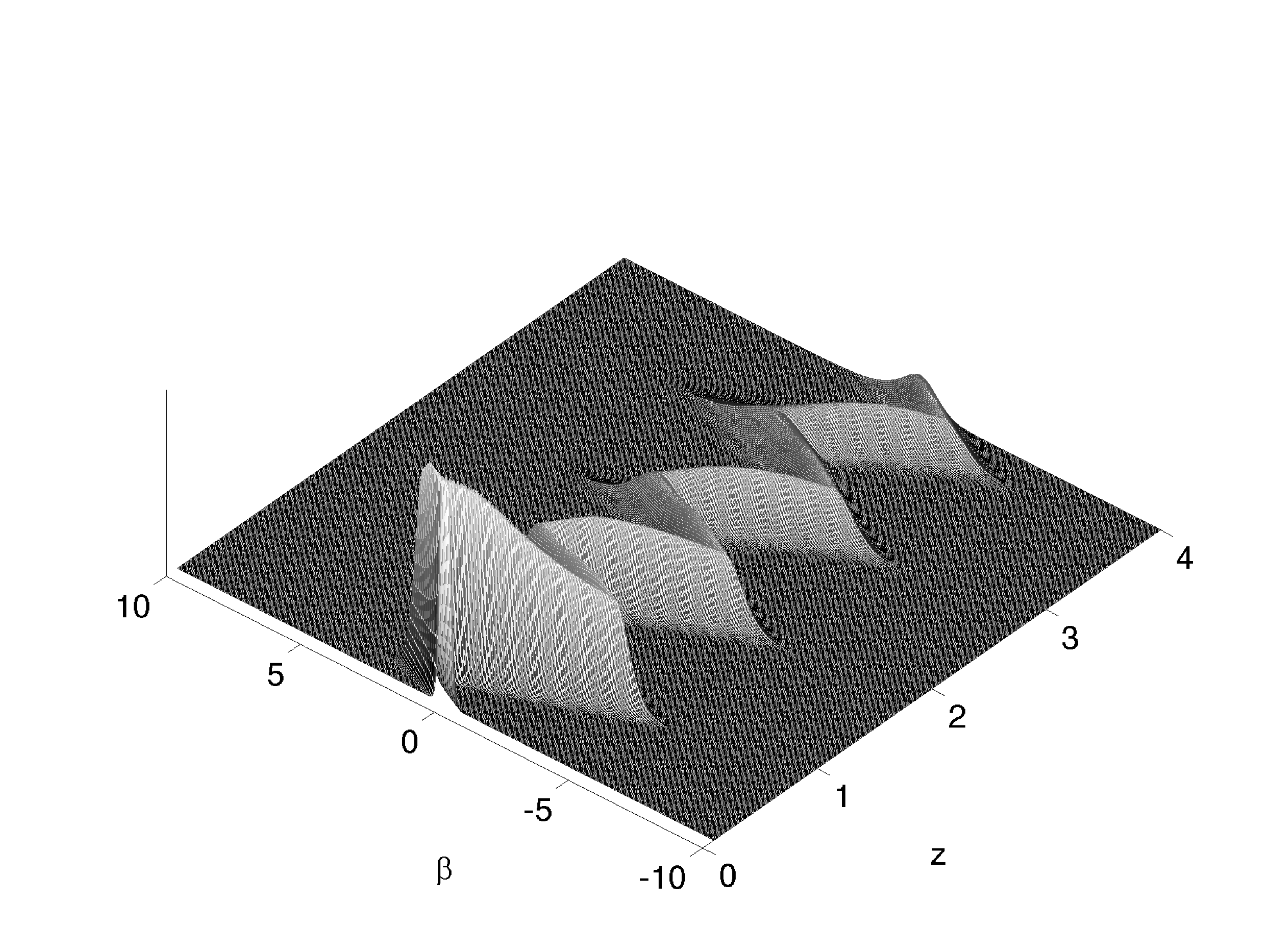}
\caption{{\small Evolution of the probability density of $\beta$}}
\label{fig:betmargin}
\end{center}
\end{minipage}
\end{figure}

Intuitively, we expect the solutions of the stochastic dynamical
system presented by (\ref{AAeqns}) to experience scale-separation as
in the deterministic case. This separation of scales should result
in fast variations of the probability densities for $\beta$ and
$\Omega$ within a span and a slow evolution of both quantities over
many periods visible in the corresponding Poincar{\'e} sections.
Numerical simulations confirm this intuition. Fig.
{\ref{fig:Ommargin}} and fig. {\ref{fig:betmargin}} present the
evolution of both probability densities.

We start our multi-scale analysis of the system by immediately
writing Eqns.~\ref{AAeqns} in the form of Eqn.~\ref{SODE}, with
$x=(\Omega\;\beta)^T$ and $t=z$.  Thus,
\begin{eqnarray}
f(x,z;\epsilon) & = & \bp 0\cr d_0(z)\Omega\ep + \epsilon\bp
3\partial h/\partial\beta\cr -\partial
h/\partial\Omega\ep,\quad\mbox{and}\label{feqn}\\
g(x,z) & = & \bp 0 & 0\cr 0 & \Omega\ep,\label{geqn}
\end{eqnarray}
where we write $\xi(z) = (\xi_1(z)\;\xi_2(z))^T$ formally even though
only $\xi_2$ is relevant.  The initial condition is simply
$a=(\Omega_0\; 0)^T$. Note that the behavior at leading order is easily solved
to yield
\begin{equation}
x^{(0)} = \bp 1&0\cr R(z_0)&1\ep y(z_1)\qquad\mbox{and} \qquad
R(z_0) = \int_0^{z_0} d_0(\zeta)\,d\zeta.
\end{equation}

In order to obtain coarse-grained equations from the original system
given by Eqn.~\ref{AAeqns}, we first express the transition
probability in the form of a path integral, i.e.,
\begin{equation}
p(x,z) = \int_{{\mathcal{C}}[\beta,z|\beta_0,0]}
{\mathcal{D}}\beta(\zeta)
\;{\mathrm{e}}^{-\frac{1}{2\epsilon}\int_0^z \frac{1}{\Omega^2}
\left(\dot \beta - d_0(\zeta)\Omega+\epsilon  \frac{\partial
h}{\partial \Omega} \right)^2 d\zeta}\;
\delta\left(\Omega(z)-\Omega_0-\int_0^z 3 \epsilon  \frac{\partial
h}{\partial \beta} d\zeta\right),
\end{equation}
with the function $h$ given by Eqn.~\ref{tm_hamiltonian}. A short
derivation of this representation is found in the appendix. For the
short-time propagator we note in particular that, since $z_{0k}=k$,
\begin{eqnarray}
p(x_k,k+1|x_k,k) &=& \int_{{\mathcal{C}}[\beta_{k+1},k+1|\beta_k,k]}
{\mathcal{D}}\beta(\zeta)
\;{\mathrm{e}}^{-\frac{1}{2\epsilon}\int_k^{k+1} \frac{1}{\Omega^2}
\left(\dot \beta - d_0(\zeta)\Omega(\zeta)+\epsilon  \frac{\partial h}{\partial \Omega} \right)^2 d\zeta}\; \nonumber \\
&\times& \delta\left(\Omega(k+1)-\Omega(k)-\int_{k}^{k+1} 3 \epsilon
\frac{\partial h}{\partial \beta} d\zeta\right).
\label{pathint_beta}
\end{eqnarray}
We wish to apply a semi-classical approximation to this short-term
propagator by finding the minimal trajectory using the associated
Euler-Lagrange equations \cite{chaichian-demichev:2001}. For this
purpose, we consider a multi-scale expansion
\begin{eqnarray}
\Omega = \Omega^{(0)}(z_0,z_1)+\epsilon \Omega^{(1)}(z_0,z_1) + \hdots \\
\beta = \beta^{(0)}(z_0,z_1)+\epsilon \beta^{(1)}(z_0,z_1) + \hdots
\end{eqnarray}
and find at the leading order
\begin{equation}
\Omega^{(0)}(z_0,z_1) = \bar \Omega(z_1)\qquad\mbox{and} \qquad
\beta^{(0)}(z_0,z_1) = R(z_0) \bar \Omega(z_1) + \bar \beta(z_1)\,.
\end{equation}
Therefore, within one period, $\Omega^{(0)}$ can be considered to be
constant and the constraint $\dot \Omega = 3\epsilon \partial
h/\partial \beta$ yields an evolution equation for the $z_1$
dependence
\begin{equation}
\bar\Omega' \equiv \frac{d\bar\Omega}{dz_1} =
\int_{k}^{k+1}3\frac{\partial h}{\partial \beta} dz_0
\end{equation}
as well as the solution of the first-order correction $\Omega^{(1)}$
\begin{equation} \label{eq_Omega_bar}
\Omega^{(1)} (z_0,z_1)= \int_{k}^{z_{0}} 3 \frac{\partial
h}{\partial \beta} d\tilde z - \bar\Omega'(z_1)z_0\,.
\end{equation}
Here, we have applied standard deterministic multi-scale analysis -
consistent with the fact that the evolution equation of $\Omega$ does
not contain any further source of noise. We can now rewrite the
Lagrangian in the kernel of the above path integral given in
Eqn.~\ref{pathint_beta} by noting that
\begin{eqnarray*}
\dot \beta - d_0(\zeta)\Omega(\zeta) - \epsilon\frac{\partial
h}{\partial \Omega} &=&
\epsilon\left(R(\zeta)\bar\Omega'+\bar\beta'+\frac{\partial\beta^{(1)}}{\partial
\zeta}
- d_0(\zeta)\Omega^{(1)}-\frac{\partial h}{\partial \Omega} \right) \\
&=& \epsilon\left(
R_0(\zeta)\bar\Omega'+\bar\beta'+\frac{\partial\beta^{(1)}}{\partial
\zeta} -d_0(\zeta) \int_{k}^{\zeta} 3 \frac{\partial h}{\partial
\beta} d\tilde z +d_0(\zeta)\bar\Omega'\zeta +\frac{\partial
h}{\partial \Omega} \right)
\end{eqnarray*}
In this way, the Lagrangian depends on $\beta^{(1)}(\zeta)$ and
$\beta^{(1)}_\zeta\equiv\partial \beta^{(1)}/
\partial \zeta$, or more formally written as
\begin{displaymath}
L=L(\beta^{(1)},\beta^{(1)}_\zeta,\zeta)
\end{displaymath}
and the corresponding Euler-Lagrange equation is given by
\begin{equation}
\frac{d}{d\zeta}\frac{\partial L}{\partial
\beta^{(1)}_\zeta}-\frac{\partial L}{\partial \beta^{(1)}} = 0
\end{equation}
Since $\beta^{(1)}$ is cyclic, we immediately find
\begin{equation}
R\bar\Omega'+\bar\beta'+\frac{\partial \beta^{(1)}}{\partial \zeta}
- d(\zeta)\int_{k}^{\zeta} 3 \frac{\partial h}{\partial \beta}
d\tilde z + d(\zeta)\bar\Omega'\zeta + \frac{\partial h}{\partial
\Omega} = C
\end{equation}
The value of the constant $C$ is found by the boundary conditions on
$\beta^{(1)}$ and for $\beta^{(1)}(k+1,z_1)=\beta^{(1)}(k,z_1)$ we
find by integration of the above equation over the fast variable $s$
immediately
\begin{equation}
C = \bar\beta' +\int_0^1 \left(3\frac{\partial h}{\partial
\beta}+\frac{\partial h}{\partial \Omega}\right)\, d\zeta
\end{equation}
which finally lets the averaged Lagrangian become
\begin{equation}
\bar L = \frac{1}{2}\frac{1}{{\bar \Omega}^2} \left[\bar\beta' +
\int_{k}^{k+1} \left(3\frac{\partial h}{\partial
\beta}+\frac{\partial h}{\partial \Omega}\right)\, d\zeta \right]^2
\end{equation}
This, however corresponds again exactly to a stochastic equation with
a noise process $\xi=\xi(z_1)$ on the slow scale given by
\begin{equation}
\bar \beta ' = \bar \Omega \xi(z_1) - \int_{0}^{1}
\left(3\frac{\partial h}{\partial \beta}+\frac{\partial h}{\partial
\Omega}\right)\, d\zeta
\end{equation}
and together with Eqn.~\ref{eq_Omega_bar}, we have found the coarse-grained system of stochastic
differential equations. Note that in the present case, it is possible to solve all integrals
analytically such that we can rewrite the system in terms of the slow variable $y$ with
$y_1 \equiv \bar \Omega$ and $y_2 \equiv \bar \beta$ as
\begin{equation} \label{ms_final_eqs}
\frac\partial{\partial z_1}\bp y_1\cr y_2\ep = \bp
\Gamma_1(y_1,y_2)\cr \Gamma_2(y_1,y_2)\ep + \bp 0\cr
y_1\Xi(z_1)\ep.
\end{equation}
The slow-scale white noise process $\Xi$ is simply given by
\begin{equation}
\Xi(z_1) = \bp 0\cr \int_0^1\xi(z_0)\,dz_0\ep = \bp
0\cr\xi(z_1)\ep. \label{MSXi}
\end{equation}
The functions $\Gamma_1$ and $\Gamma_2$ can be expressed as
\begin{eqnarray}
\Gamma_1(y) &=& \frac6{\hat{d}y_1}\left[\frac{\partial
H}{\partial\beta}(y_1,y_2+\frac14\hat{d}y_1)
- \frac{\partial H}{\partial\beta}(y_1,y_2-\frac14\hat{d}y_1)\right] \\
&=& \frac{4C_2\sqrt{y_1}}{\hat{d}}\left\{
\left[C_1+(y_2+\frac14\hat{d}y_1)^2\right]^{-1/2} -
\left[C_1+(y_2-\frac14\hat{d}y_1)^2\right]\right\}\nonumber\\
\end{eqnarray}
and
\begin{eqnarray}
\Gamma_2(y) &=&
\frac3{\hat{d}y_1^2}\left[H(y_1,y_2+\frac14\hat{d}y_1) -
H(y_1,y_2-\frac14\hat{d}y_1)\right] \nonumber\\
&&- \frac3{2y_1}\left[ \frac{\partial
H}{\partial\beta}(y_1,y_2+\frac14\hat{d}y_1) + \frac{\partial
H}{\partial\beta}(y_1,y_2-\frac14\hat{d}y_1)\right]
+\dav y_1\\
 &=& \frac{2C_2}{\hat{d}\sqrt{y_1}}
 \left[\sinh^{-1}\left(\frac{y_2+\frac14\hat{d}y_1}{\sqrt{C_1}}\right)
 -
 \sinh^{-1}\left(\frac{y_2-\frac14\hat{d}y_1}{\sqrt{C_1}}\right)\right]
 \nonumber\\
 &&-C_2\sqrt{y_1} \left\{ \left[ C_1+(y_2+\frac14\hat{d}y_1)^2
 \right]^{-1/2} + \left[ C_1 +
 (y_2-\frac14\hat{d}y_1)^2\right]^{-1/2}\right\} +\dav
 y_1.\nonumber\\
\end{eqnarray}
with
\begin{equation}
H(\Omega,\beta) \equiv
\frac23C_2\Omega^{3/2}\sinh^{-1}(\beta/\sqrt{C_1}),
\end{equation}
Although the above analysis was carried out for a particular example, it shows the
steps that are necessary to apply and modify the method for other cases.
Note that it is not a requirement that the occurring integrals
can be solved analytically. Moreover, using the semi-classical method in order to
calculate the short-term propagator is appropriate in this case since the evolution
of the system is linear on a short time scale. For systems that show nonlinear behavior
on short times, other methods of evaluating the corresponding path integral can be applied.

\section{Perturbative analysis of Fokker-Planck equation}
\label{FPsec}

In this section we show that a perturbative analysis of the
corresponding Fokker-Planck equation will yield for the particular
example exactly the same result as the path integral based method.
The idea to perform asymptotic analysis on the level of the Fokker-Planck
equation in order to characterize the behavior of the associated
stochastic process has been applied very successfully in a variety of
contexts \cite{papanicolaou:1977}. Since the Fokker-Planck equation
is deterministic, we can use the classic theory of deterministic
multi-scale expansions, adapted to the Fokker-Planck equation.

Hence, instead of considering the path integral representation of
the stochastic process $x(t)$, we now are considering the evolution
equation of the probability density $p(x,t)$ given by the Fokker-Planck
equation
\begin{equation}
\frac{\partial p}{\partial t} = -\frac{\partial}{\partial x_i}(f_ip)
+ \frac12\epsilon\frac{\partial}{\partial x_i}
\left(g_{ik}\frac{\partial}{\partial x_j}(g_{jk}p)\right),
\label{FPeqn}
\end{equation}
with initial condition $p(x,0) = \prod_{i=1}^n\delta(x_i-a_i)$. Note
that, as before, we are using Einstein's summation convention and
the Stratonovich interpretation of Eqn.~\ref{SODE}
\cite{gardiner:1985}.

Again, the secular growth we obtain when trying a naive perturbation expansion
for $p$ suggests we adopt a multiple-scales approach:
\begin{equation}
p(x,t) = p^{(0)}(x,t_0,t_1,\dots) + \epsilon
p^{(1)}(x,t_0,t_1,\dots) + \dots,
\end{equation}
where $t_j=\epsilon^jt$.  In this case, we have no ambiguity in the
scale at which the effect of randomness is felt, since the
Fokker-Planck equation is completely deterministic. The probability
density $p$ captures all moments of $x(t)$. The leading order of
this expansion satisfies
\begin{equation}
\frac{\partial p^{(0)}}{\partial t_0} = -\frac{\partial}{\partial
x_i}\left[f_i(x,t_0;0) p^{(0)}\right],
\end{equation}
so that along characteristics $x=x_c(t_0;y)$ where
\begin{equation}
\frac{dx_c}{dt_0} = f(x_c(t_0),t_0;0)\quad\mbox{and}\quad x_c(0)=y,
\end{equation}
we have that
\begin{equation}
\frac{dp^{(0)}}{dt_0} = -\left[\frac{\partial}{\partial x_i}
f_i(x_c(t_0),t_0;0)\right]p^{(0)}.
\end{equation}
The $t_0$ dependence has been written as a full derivative here to
denote the fact that it is taken along characteristics; naturally, $p$
also depends on the slower time scale $t_1$.
The solution is then
\begin{equation} p^{(0)} =
P(y,t_1,\dots)\exp\int_0^{t_0}-\frac{\partial}{\partial x_i}
f_i(x_c(\zeta),\zeta;0)\,d\zeta
\end{equation}
with initial condition
\begin{equation}
P(y,0) = \prod_{i=1}^n \delta(y_i-a_i).
\end{equation}
The next order is
\begin{equation}
Lp^{(1)} = -\frac{\partial p^{(0)}}{\partial t_1} -
\frac{\partial}{\partial x_i}\left(\left.\frac{\partial
f_i}{\partial \epsilon}\right|_{\epsilon=0}p^{(0)}\right) + \frac12
\frac\partial{\partial x_i}\left(g_{ik}\frac\partial{\partial
x_j}(g_{jk}p^{(0)})\right) \label{FPordereps}
\end{equation}
where
\begin{equation}
L\equiv \frac{\partial}{\partial t_0} +
\frac{\partial}{\partial_i}\left[f_i(x,t;0)\circ\right],
\end{equation}
with adjoint operator
\begin{equation}
L^\dag = -\frac{\partial}{\partial
t_0}-f_i(x,t;0)\frac{\partial}{\partial x_i}.
\end{equation}
By comparison with the leading order, it is clear that $\ker
L^\dag$ is spanned by functions that are constant in $t_0$ along
characteristics $x_c(t_0;y)$, i.e., any square-integrable function
$\phi(y)$ (in fact, we have thus far omitted a discussion of the
appropriate boundary
conditions, but for simplicity we will assume
$t^*$-periodicity in $t_0$ and sufficiently fast decay in $|x|$).
The Fredholm alternative therefore requires that
\begin{eqnarray}
&0 = &\int_0^{t^*}\,dt_0\int\,d^nx\,\phi(y)\left[-\frac{\partial
P}{\partial t_1} \exp\int_0^{t_0}-\frac\partial{\partial x_i}
f_i(x_c(\zeta),\zeta;0)\,d\zeta
\right.\nonumber\\
&&\left. -\frac\partial{\partial x_i}\left(\left.\frac{\partial
f_i}{\partial\epsilon}\right|_{\epsilon=0}
P(y,t_1)\exp\int_0^{t_0}-\frac\partial{\partial x_j}
f_j(x_c(\zeta),\zeta;0)\,d\zeta\right)
\right.\nonumber\\
&&\left. +\frac12\frac\partial{\partial x_i}\left(
g_{ik}\frac\partial{\partial x_j}(g_{jk}P(y,t_1))\right)
\exp\int_0^{t_0}-\frac\partial{\partial x_l} f_l(x_c(\zeta),\zeta;0)\,d\zeta \right]\nonumber\\
\label{FPfred}
\end{eqnarray}
for all members $\phi(y)$ of the admissible class of functions. This
clearly leads to an IBVP for $P(y,t_1)$ which is now, depending on the
nonempty elements of the matrix $g(x,t)$,
parabolic rather than hyperbolic.  The particular form of this
equation is seen most easily when the characteristic coordinate
transformation $x=x_c(t_0;y)$ is invertible, as we can then write
down directly the transformed equation in $y$ and $t_1$.  This
will be the case for our current example, as we will see below.

In application the concrete problem given by Eqns.~\ref{AAeqns},
we note immediately from Eqn.~\ref{feqn} that
\begin{equation}
\frac\partial{\partial x_i} f_i(x,z;0) = 0,
\end{equation}
which greatly simplifies matters.  As before, the leading order
characteristics are given by
\begin{equation}
x_c(z_0) = \bp 1&0\cr R(z_0)&1\ep y,
\end{equation}
with the leading-order solution simply given by
\begin{equation}
p^{(0)}(x,z_0,z_1,\dots) = P(y,z_1,\dots).
\end{equation}
After using the above characteristics to change variables from $x$
to $y$, the dependence of $P$ on $y$ and $z_1$, as imposed by the
Fredholm alternative condition in Eqn.~\ref{FPfred}, is
given by
\begin{eqnarray}
\frac{\partial P}{\partial z_1} &=& \int_0^1dz_0 \left\{
-3\left(\frac\partial{\partial
  y_1}-R(z_0)\frac\partial{\partial y_2}\right) \left[\frac{\partial
    h}{\partial\beta}(y_1,y_2+R(z_0)y_1)P(y,z_1)\right] \right.\nonumber\\
&& \left.+\frac\partial{\partial y_2}\left[\frac{\partial
    h}{\partial\Omega}(y_1,y_2+R(z_0)y_1) P(y,z_1)\right] +
\frac12y_1^2\frac{\partial^2P}{\partial y_2^2}\right\}\nonumber\\
&=& -\dav y_1\frac{\partial P}{\partial y_2}
-3\frac\partial{\partial
y_1}\int_0^1\frac{\partial^2H}{\partial\beta^2}(\cdot,\cdot)P(y,z_1)\,dz_0
\nonumber\\ &&+ \frac\partial{\partial
y_2}\int_0^1\left[\frac{\partial^2H}{\partial\Omega\partial\beta}(\cdot,\cdot)
+3R(z_0)\frac{\partial^2H}{\partial\beta^2}(\cdot,\cdot)\right]P(y,z_1)\,dz_0+\frac12 y_1^2\frac{\partial^2P}{\partial y_2^2},
\end{eqnarray}
where the derivatives of $H(\Omega,\beta)$ are evaluated along the
characteristics, that can be written explicitly as $\Omega=y_1$,
$\beta = y_2+R(z_0)y_1$. Finally, after some algebra we observe that
the above evaluates to
\begin{equation} \label{fp_final_eq}
\frac{\partial P}{\partial z_1} = -\frac\partial{\partial
y_1}(\Gamma_1P) - \frac\partial{\partial y_2}(\Gamma_2P) +
\frac12y_1^2\frac{\partial^2P}{\partial y_2^2},
\end{equation}
which is exactly the Fokker-Planck equation for the slow process
given by Eq.~\ref{ms_final_eqs} obtained in Sec.~\ref{Dispsec}.

\section{Conclusion}
\label{Concsec}

We have presented a new method to derive equations that describe the
slow evolution of a system of stochastic ordinary differential
equations.  The conceptual basis of the method is the expression of
state transition probabilities as path integrals, where a separation
of scales allows a decomposition of the paths into self-similar
segments.  By performing a multi-scale expansion of the Lagrangian
in each segment, we obtain an effective path integral that
represents the coarse-grained noise process relevant to the slow
dynamics of the system.  From the appropriate limit of this path
integral (i.e., refinement of the path partition), we obtain a new
system of stochastic differential equations that describe behavior
over long time scales.  We have applied our method to a stochastic
system arising in nonlinear optics, and have shown that our method
produces a leading-order result that is identical to that obtained
from a standard asymptotic analysis of the singularly perturbed
Fokker-Planck equation associated with the stochastic system. The
conditions necessary for this agreement to hold is a topic of
ongoing research.

\section*{Acknowledgments}

The work of T. Sch{\"a}fer was supported by CUNY Research Foundation
through the grant PSCREG-38-860 and through the Center of
Interdisciplinary Applied Mathematics and Computational Sciences at
the College of Staten Island. R. O. Moore gratefully acknowledges
support from the NSF through grant NSF-DMS 0511091.

\newpage

\appendix

\section{Path integral representation}

In this appendix we give a brief overview of how to derive a path
integral representation of a stochastic system following
\cite{langouche-roekaerts-etal:1982}. We start with a stochastic
equation of the form (\ref{SODE}),
\begin{equation} \label{sode_2}
\dot{x} = f(x,t;\epsilon) + \sqrt\eta g(x,t)\xi(t),\quad x(0)=a,
\end{equation}
where we have denoted the noise strength by $\eta$ instead of
$\epsilon$ to reflect the fact that noise need not be small in this
formalism. We take (\ref{sode_2}) to be the limit of difference
equations for $\Delta x_i^{(k)}$ (where the subscript and
superscript now represent the component index and the time
increment, respectively) where
\begin{equation}
\Delta x_i^{(k)} = f_i^{(k)}\Delta t + \sqrt\eta\left(g_{ij}^{(k)} +
s\frac{\partial g_{ij}^{(k)}}{\partial x_l}\Delta x_l^{(k)}\right)
\Delta w_j^{(k)}.
\end{equation}
The time increments $\Delta t$ identify a regular partition of
interval $[0,t]$ with $K$ intermediate points.  All terms in the
stochastic difference equation are evaluated at the {\em pre-point},
hence
\begin{displaymath}
f_i^{(k)}=f_i(x^{(k-1)},t^{(k-1)}), \qquad
g_{ij}^{(k)}=g_{ij}(x^{(k-1)},t^{(k-1)}),
\end{displaymath}
and the parameter $s$ accounts for the different possible
definitions of the stochastic integral: a choice of $s=0$
corresponds to the Ito interpretation of (\ref{sode_2}) and a choice
of $s=1/2$ corresponds to the Stratonovich interpretation.  We can
rewrite $\Delta x_i^{(k)}$ using the fact that the Brownian
increments $\Delta w_i^{(k)}$ are uncorrelated and of order
$\sqrt{\Delta t}$. This yields
\begin{equation}
\Delta x_i^{(k)} = \left(f_i^{(k)} + \eta s \frac{\partial
g_{ij}^{(k)}}{\partial x_l}g_{lj}^{(k)}\right)\Delta t +\sqrt\eta
g_{ij}^{(k)}\Delta w_j^{(k)}
\end{equation}
Defining now the inverse of $g$ to be $h=g^{-1}$, we can solve for
the $\Delta w_i^{(k)}$ and obtain
\begin{equation} \label{sol_w}
\Delta w_i^{(k)} = \frac{h_{ij}^{(k)}}{\sqrt\eta}\left( \Delta
x_j^{(k)} - \left( f_j^{(k)} + \eta s \frac{\partial
g_{ij}^{(k)}}{\partial x_l}g_{lj}^{(k)}\right)\Delta t\right)
\end{equation}
Note that if $g$ is not invertible, an appropriate regularization
has to be chosen (see Appendix B). In this section, however, we will
simply take $g$ to be nonsingular for all times $t$ and all points
$x$.

The joint probability distribution of a path $(w)\equiv
(w^{(1)},...,w^{(K+1)})$ is given by
\begin{equation}
p((w))=\prod_{k=1}^{K+1}\frac{1}{\sqrt{2\pi\Delta t}}\prod_{i=1}^n
\Delta w_i^{(k)}{\mathrm{e}}^{-\frac{1}{2\Delta t} (\Delta
w_i^{(k)})^2}.
\end{equation}
The derivation of a path integral representation for the stochastic
differential equation (\ref{sode_2}) amounts to performing a
coordinate transformation in the above probability density from the
path $(w)$ to the path $(x)$ given by
$(x)\equiv(x^{(1)},...,x^{(K+1)})$. The determinant $|J|$ of the
Jacobian of this transformation is simple, since the discretization
is based on the pre-point. We find
\begin{equation}
|J| = \prod_{k=1}^{K+1}\left|\begin{array}{ccc}
\frac{\partial w_{1}^{(k)}}{\partial x_1^{(k)}} & \hdots & \frac{\partial w_{1}^{(k)}}{\partial x_n^{(k)}} \\
\vdots & \vdots & \vdots \\
\frac{\partial w_{n}^{(k)}}{\partial x_1^{(k)}} & \hdots &
\frac{\partial w_{n}^{(k)}}{\partial x_n^{(k)}}
\end{array}\right| =
\prod_{k=1}^{K+1}\eta^{-n/2}{\mathrm{det}}([h_{ij}^{(k)}]).
\end{equation}
We now define the action $S$ as the following limit for
$K\rightarrow\infty$ and $\Delta t \rightarrow 0$:
\begin{eqnarray*}
S & = & \lim_{K\rightarrow\infty,t\rightarrow 0}
\frac{1}{2\Delta t}\sum_{k=1}^{K+1}\sum_{i=1}^n (\Delta w_i^{(k)})^2  \\
&=& \lim_{K\rightarrow\infty,t\rightarrow 0} \frac{1}{2\Delta t}
\sum_{k=1}^{K+1}\sum_{i=1}^n \left(
\frac{h_{iq}^{(m)}}{\sqrt\eta}\left( \Delta x_q^{(k)} - \left(
f_q^{(k)} + \eta s \frac{\partial g_{qj}^{(k)}}{\partial
x_l}g_{lj}^{(k)}\right)\Delta t\right)
 \right)^2 \\
&=& \lim_{K\rightarrow\infty,t\rightarrow 0}
\frac{1}{2\eta}\sum_{k=1}^{K+1}\sum_{i=1}^n
\left(h_{iq}^{(k)}\right)^2 \left(\frac{\Delta x_q^{(k)}}{\Delta t}
- \left(f_q^{(k)} + \eta s \frac{\partial g_{qj}^{(k)}}{\partial
x_l}g_{lj}^{(k)}\right) \right) ^2 \Delta t\\
%
%
 &=& \frac{1}{2\eta} \int_{0}^{t} \sum_{i=1}^{n} h_{ik}^2
\left(\dot x_k -\left(f_k+\eta s \frac{\partial g_{kj}}{\partial
x_l}g_{lj}\right)\right)^2\,d\tau,
\end{eqnarray*}
where we have suppressed the arguments $(x(\tau),\tau)$ in $h_{ik}$,
$f_k$, etc., for simplicity of presentation.  Defining
\begin{eqnarray*}
{\mathcal{D}}x(\tau) &=& \lim_{K\rightarrow\infty,t\rightarrow 0}
\prod_{k=1}^{K+1}\frac{1}{\left(\sqrt{2\pi\Delta t}\right)^n}
\frac{1}{\eta^{n/2}}\,{\mathrm{det}}( [h_{ij}^{(k)}] )\,
\prod_{i=1}^n dx_i^{(k)} \\
&\equiv & \prod_{\tau = 0}^{t} \frac{1}{\left(\sqrt{2\pi\,
d\tau}\right)^n} {\mathrm{det}}( [h_{ij} (x(\tau),\tau) ])
\prod_{i=1}^n dx_i(\tau)
\end{eqnarray*}
we can write the path integral representation for the transition
probability from the initial condition $x(0)=a$ to the point $x(t)=x$
as
\begin{equation}
p(x,t) = \int_{{\mathcal{C}}(x,t|a,0)} {\mathcal{D}}x(\tau)
\;{\mathrm{e}}^{-L(x(\tau),\dot x(\tau),\tau)},
\end{equation}
where ${\mathcal{C}}(x,t|a,0)$ denotes the set of all possible
paths that connect the point $(a,0)$ to the point $(x,t)$.

\section{Regularized path integral representation \\
of the $(\boldsymbol\Omega,\boldsymbol\beta)$ equations}
In this section we show how to derive a path integral representation
in $(\Omega,\beta)$ for the particular system given by
Eqns.~(\ref{AAeqns}). For simplicity we rewrite these equations as
\begin{eqnarray}
\frac{d\Omega}{dz} & = & f_1(\Omega,\beta,z)+\kappa \sqrt\eta \,\xi_1(z)
\label{Omega_beta_sde_1} \\
\frac{d\beta}{dz} &=& f_2(\Omega,\beta,z) + \Omega \sqrt\eta\,\xi_2(z)
\label{Omega_beta_sde_2}
\end{eqnarray}
where we obtain (\ref{AAeqns}) for $\kappa = 0$. Similar
regularization techniques are well-established, e.g., in the case of
Brownian motion \cite{chaichian-demichev:2001}. The matrices $g$ and
$h$ in (\ref{sol_w}) are given by
\begin{equation}
g = \left(\begin{array}{cc} \kappa & 0 \\ 0 & \Omega \end{array}
\right), \qquad h = \left(\begin{array}{cc} \kappa^{-1} & 0 \\ 0 &
\Omega^{-1} \end{array} \right)
\end{equation}
such that
\begin{equation}
h_{ik}\frac{\partial g_{kj}}{\partial x_l}g_{lj}=0
\end{equation}
for $i=1,2$. With $x=(\Omega,\beta)^T$ and $a
=(\Omega_0,\beta_0)^T$, we can write the corresponding path integral
representation of (\ref{Omega_beta_sde_1}, \ref{Omega_beta_sde_2})
as
\begin{equation}
p(x,z) = \int_{{\mathcal{C}}[x,z|a,0]}
\prod_{\zeta=0}^{z}\frac{d\Omega(\zeta)}{\sqrt{2\pi d\zeta}}
\frac{d\beta(\zeta)}{\sqrt{2\pi
d\zeta}}\frac{1}{\kappa\sqrt{\eta}}\frac{1}{\Omega\sqrt{\eta}}\;
\;{\mathrm{e}}^{-\frac{1}{2\eta}\int_0^z \frac{1}{\kappa^2}(\dot
\Omega - f_1)^2 d\zeta}\; \;{\mathrm{e}}^{-\frac{1}{2\eta}\int_0^z
\frac{1}{\Omega^2}(\dot \beta - f_2)^2 d\zeta}.
\end{equation}
In order to take the limit $\kappa \rightarrow 0$, we note that
\begin{equation}
\lim_{\kappa\rightarrow 0} \prod_{\zeta=0}^{z} \frac{1}{\sqrt{2\pi
d\zeta}} \frac{1}{\kappa\sqrt{\eta}}
\;{\mathrm{e}}^{-\frac{1}{2\eta}\int_0^z \frac{1}{\kappa^2}(\dot
\Omega - f_1)^2 d\zeta} =  \prod_{\zeta=0}^{z}
\frac{1}{d\zeta}\delta(\dot \Omega - f_1),
\end{equation}
as can been seen by straightforward calculations using the same
discretization as in the derivation of the path integral
representation. Integration over $d\Omega(\zeta)$ yields then
\begin{eqnarray*}
p(x,z) &=& \int_{{\mathcal{C}}[\beta,z|\beta_0,0]}
\prod_{\zeta=0}^{z} \frac{d\beta(\zeta)}{\sqrt{2\pi d\zeta}}
\frac{1}{\Omega\sqrt{\eta}}\;
\;{\mathrm{e}}^{-\frac{1}{2\eta}\int_0^z \frac{1}{\Omega^2}(\dot
\beta - f_2)^2 d\zeta}\;
\\
&\times& \delta\left(\Omega(z)-\Omega_0-\int_0^z
f_1(\Omega(\zeta),\beta(\zeta),\zeta)d\zeta\right)
\end{eqnarray*}
or, written in a more compact way,
\begin{equation}
p(x,z) = \int_{{\mathcal{C}}[\beta,z|\beta_0,0]}
{\mathcal{D}}\beta(\zeta) \;{\mathrm{e}}^{-\frac{1}{2\eta}\int_0^z
\frac{1}{\Omega^2}(\dot \beta - f_2)^2 d\zeta}\;
\delta\left(\Omega(z)-\Omega_0-\int_0^z
f_1(\Omega(\zeta),\beta(\zeta),\zeta)d\zeta\right).
\end{equation}

\bibliography{master}

\end{document}